\documentclass{PoS}

\newcommand{\kslash}{k\kern-1ex /}
\newcommand{\pslash}{p\kern-1ex /}
\newcommand{\qslash}{q\kern-1ex /}
\newcommand{\lslash}{l\kern-1ex /}
\newcommand{\sslash}{s\kern-1ex /}
\newcommand{\Dslash}{D\kern-1.2ex /}

\newcommand{\beqa}{\begin{eqnarray}}
\newcommand{\eeqa}{\end{eqnarray}}

\newcommand{\bd}{\begin{description}}
\newcommand{\ed}{\end{description}}

\newcommand{\ben}{\begin{eqnarray}}
\newcommand{\een}{\end{eqnarray}}
\newcommand{\nn}{\nonumber}
\def\lsim{\raise0.3ex\hbox{$<$\kern-0.75em\raise-1.1ex\hbox{$\sim$}}}
\def\gsim{\raise0.3ex\hbox{$>$\kern-0.75em\raise-1.1ex\hbox{$\sim$}}}
\def\simgt{\rlap{\lower 6.0 pt\hbox{$\mathchar \sim$}}\raise 2.5pt \hbox {$>$}}
\def\simlt{\rlap{\lower 6.0 pt\hbox{$\mathchar \sim$}}\raise 2.5pt \hbox {$<$}}

\title{$N_f=2+1$ dynamical Wilson quark simulation toward the physical point}

\ShortTitle{$N_f=2+1$ dynamical Wilson quark simulation toward the physical point}

\author{Yoshinobu Kuramashi\thanks{E-mail: 
kuramasi@het.ph.tsukuba.ac.jp}\hspace{3mm}for the PACS-CS Collabolation \\
Center for Computational Sciences and
Graduate School of Pure and Applied Sciences,\\ 
University of Tsukuba, Tsukuba, Ibaraki 305-8571, Japan}
\abstract{
We present preliminary results of the
PACS-CS project which simulates 2+1
flavor lattice QCD toward the physical point 
with the nonperturbatively $O(a)$-improved
Wilson quark action and the Iwasaki gauge action.
Calculations are carried out at $\beta=1.9$ on a
$32^3\times 64$ lattice with the use of the domain-decomposed HMC algorithm
to reduce the up-down quark mass.
The resulting pseudoscalar meson masses range from 730~MeV down to 210~MeV.
We discuss the physical results including the chiral analysis in
the pseudoscalar meson sector and the hadron spectrum.
Some algorithmic issues are also discussed.
}

\FullConference{The XXV International Symposium on Lattice Field Theory\\
		 July 30-4 August 2007\\
		 Regensburg, Germany}

\begin{document}

\section{Introduction}

The first lattice QCD calculation of hadron masses in 1981
revealed its potential ability to nonperturbatively evaluate the
physical quantities in the strong interaction from  first principles.
Since then, the history of lattice QCD has been a succession of enduring efforts
to control the major systematic errors due to finite lattice size,
finite lattice spacing, quenching and chiral extrapolation.  
Thanks to recent progress of simulation algorithms and increasing
availability of computational resources, we are about to
bring all the above systematic errors under control. 
This will allow us to establish whether or not QCD is the fundamental theory of the strong
interaction by investigating the hadron spectrum, and further proceed to elucidate 
the fundamental issues of the strong interactions and the Standard Model.
 
The previous CP-PACS/JLQCD project\cite{cppacs/jlqcd1,
cppacs/jlqcd2} aimed at a full removal of the quenching effects 
by performing $N_f=2+1$ lattice QCD simulations 
with the nonperturbatively $O(a)$-improved Wilson quark action\cite{csw}
and the Iwasaki gauge action\cite{iwasaki} 
on a $(2\rm ~fm)^3$ lattice at three lattice spacings.
While we have succeeded in incorporating the dynamical strange quark
effects by the Polynomial Hybrid Monte Carlo (PHMC) algorithm\cite{phmc},  
the lightest up-down quark mass reached with the HMC algorithm was 
64~MeV  corresponding to $m_{\pi}/m_{\rho}\approx0.6$, which required
a long chiral extrapolation to the physical point at $m_{\pi}/m_{\rho}\approx0.18$.
  
The PACS-CS(Parallel Array Computer System for Computational
Science) project\cite{ukawa1, ukawa2, kura_lat06}
is the successor to the CP-PACS/JLQCD project, which 
takes up the task that the latter has left off, namely simulation at the
physical point to remove the ambiguity of chiral extrapolation.  
It employs the same quark and gauge actions 
as the CP-PACS/JLQCD project,  
but uses the PACS-CS computer with 
a total peak speed of 14.3 TFLOPS 
developed and installed at University of Tsukuba on 1 July 2006.
The up-down quark masses are reduced by employing
the domain-decomposed HMC (DDHMC) algorithm 
with the replay trick proposed by L\"uscher\cite{luscher, kennedy}.
So far, we have reached the up-down quark mass of 6~MeV
which yields the pion mass of about 210~MeV.
We also improve the simulation of the strange quark part with 
the UV-filtered PHMC (UV-PHMC) algorithm\cite{ishikawa_lat06}.

In this report we present simulation details and preliminary results
which include the chiral analysis on 
the pseudoscalar meson masses and the decay constants with
chiral perturbation theory,
the light hadron spectrum and the $\rho$-$\pi\pi$ mixing effects.
Some algorithmic issues are also discussed. 
Selected topics on the light hadron spectrum 
and the ChPT analysis on the pseudoscalar meson masses 
and the decay constants are also reported in Refs.\cite{ukita_lat07,kadoh_lat07}.

\section{Simulation details}

\subsection{Simulation parameters}

\begin{table}[b!] 
\setlength{\tabcolsep}{10pt}
\renewcommand{\arraystretch}{1.2}
\centering
\begin{tabular}{cccccccc} \hline
$\kappa_{\rm ud}$ & $\kappa_{\rm s}$ &$\tau$& $(N_0,N_1,N_2)$ & $N_{\rm poly}$ & MD time & $\tau_{\rm int}[P]$ \\ \hline \hline
0.13700 & 0.13640 &0.5& (4,4,10) &180& 2000 & 38.2(17.3)\\
0.13727 & 0.13640 &0.5& (4,4,14) &180& 2000 & 20.9(10.2)\\
0.13754 & 0.13640 &0.5& (4,4,20) &180& 2500 & 19.2(8.6)   \\
        & 0.13660 &0.5& (4,4,28) &220& 900 & 10.3(2.9)   \\ 
0.13770 & 0.13640 &0.25& (4,4,16) &180& 2000& 38.4(25.2)\\
0.13781 & 0.13640 &0.25& (4,4,48) &180& 350 &9.1(6.1)   \\ \hline
\end{tabular}
\caption{Simulation parameters. MD time is the number of
trajectories multiplied by the trajectory length $\tau$.
$\tau_{\rm int}[P]$ denotes the integrated autocorrelation time for 
the plaquette.}
\label{tab:param}
\end{table} 

We employ the $O(a)$-improved Wilson quark action with a nonperturbative
improvement coefficient $c_{\rm SW}=1.715$\cite{csw} and 
the Iwasaki gauge action at $\beta=1.90$ on a $32^3\times64$ lattice
which is enlarged from $20^3\times40$ in the CP-PACS/JLQCD 
project to investigate the baryon masses.
Simulation parameters are summarized in Table~\ref{tab:param}.
We choose six combinations of 
the hopping parameters $(\kappa_{\rm ud}, \kappa_{\rm s})$
based on the previous CP-PACS/JLQCD results.
Among them the heaviest combination
$(\kappa_{\rm ud}, \kappa_{\rm s})=(0.13700,
0.13640)$ in this work is 
the lightest one in the previous CP-PACS/JLQCD simulations, 
which enable us to make a 
direct comparison of the two results with different lattice sizes.
As for the strange quark,  the hopping parameter $\kappa_{\rm s}=0.13640$ 
corresponds to the physical point $\kappa_{\rm s}=0.136412(50)$ as estimated 
in the CP-PACS/JLQCD work\cite{cppacs/jlqcd1, cppacs/jlqcd2}.
This is the reason why all our simulations are carried out with $\kappa_{\rm s}=0.13640$, 
the one exception being the run at $\kappa_{\rm s}=0.13660$ and $\kappa_{\rm ud}=0.13754$
to investigate the strange quark mass dependence. 
 
In order to simulate the degenerate up-down quarks
we employ the DDHMC algorithm,
whose effectiveness for small quark mass region has already
been shown in the $N_f=2$ case\cite{luscher,del06,del07}.
The characteristic feature of this algorithm is a geometric separation 
of the up-down quark determinant into the UV and the IR parts
as a preconditioner of HMC, which  
is implemented by domain-decomposing the full lattices with small blocks.
We choose $8^4$ for the block size being less than (1~fm)$^4$ in
physical units and small enough to reside within a computing node of the PACS-CS computer.
There are two prominent points in the DDHMC algorithm. Firstly,
communication between the computing nodes is not required in calculating the UV part, 
which is a preferable feature for alleviating the problem of 
a widening gap between the processor floating point performance and 
the network communication bandwidth with parallel computers.
Secondly, we can incorporate the multiple time scale 
integration scheme\cite{sexton} 
to reduce the simulation cost efficiently.
The relative magnitudes of the force terms are found to be 
\begin{eqnarray}   
||F_{\rm g}||:||F_{\rm UV}||:||F_{\rm IR}|| \approx 16:4:1,
\label{eq:force}
\end{eqnarray}
where we adopt the convention $||M||^2=-2{\rm tr}(M^2)$
for the norm of an element $M$ of the SU(3) Lie algebra.
$F_{\rm g}$ denotes the gauge part and $F_{\rm UV, IR}$ are for the UV
and the IR parts of the up-down quarks.
The associated step sizes for the forces are controlled by
three integers $N_{0,1,2}$: 
$\delta\tau_{\rm g}=\tau/N_0 N_1 N_2,\ \  \delta\tau_{\rm UV}=\tau/N_1
N_2,\ \  \delta\tau_{\rm IR}=\tau/N_2$ with $\tau$ the trajectory
length.
The integers $N_{0,1,2}$ are chosen such that 
\begin{eqnarray}
 \delta\tau_{\rm g} ||F_{\rm g}|| \approx \delta\tau_{\rm UV} ||F_{\rm UV}|| \approx \delta\tau_{\rm IR} ||F_{\rm IR}||.
\end{eqnarray} 
Taking account of the relative magnitudes of the forces in eq.(\ref{eq:force})
we find a larger value is allowed for $\delta\tau_{\rm IR}$ compared
to $\delta\tau_{\rm g}$ and $\delta\tau_{\rm UV}$,
which means that we need to calculate $F_{\rm IR}$ less frequently
in the molecular dynamics trajectories.
Since the calculation of  $F_{\rm IR}$ contains the quark matrix
inversion on the full lattice, which is the most time consuming part, 
this integration scheme 
saves the simulation cost remarkably.
The values for $N_{0,1,2}$ are listed in Table~\ref{tab:param}, where
$N_0$ and $N_1$ are fixed at 4 for all the hopping parameters, 
while the value of $N_2$ is adjusted taking account of acceptance rate 
and simulation stability.

For the UV-PHMC algorithm for the strange quark,  
the domain-decomposition is not implemented.
Since we have found $||F_{\rm s}||\approx ||F_{\rm IR}||$, 
the step size is chosen as $\delta\tau_{\rm s}=\delta\tau_{\rm IR}$.
The polynomial order for UV-PHMC, which is denoted 
by $N_{\rm poly}$ in Table~\ref{tab:param},
is adjusted to yield high acceptance rate for the global Metropolis
test at the end of each trajectory.

The inversion of the Wilson-Dirac operator $D$ on the full lattice 
is carried out by 
the SAP (Schwarz alternating procedure) preconditioned GCR solver, where 
the preconditioning can be accelerated with the single-precision
arithmetic whereas the
GCR solver is implemented with the double precision\cite{sap+gcr}. 
We employ the stopping condition $|Dx-b|/|b|<10^{-9}$ for the force
calculation and $10^{-14}$ for the Hamiltonian, which guarantees 
the reversibility of the molecular dynamics trajectories to high
precision: $|\Delta U|<10^{-12}$ for the link variables and 
$|\Delta H|<10^{-8}$ for the Hamiltonian at 
$(\kappa_{\rm ud}, \kappa_{\rm s})=(0.13781,0.13640)$.

\subsection{Plaquette history and autocorrelation time}
\begin{figure}[t!]
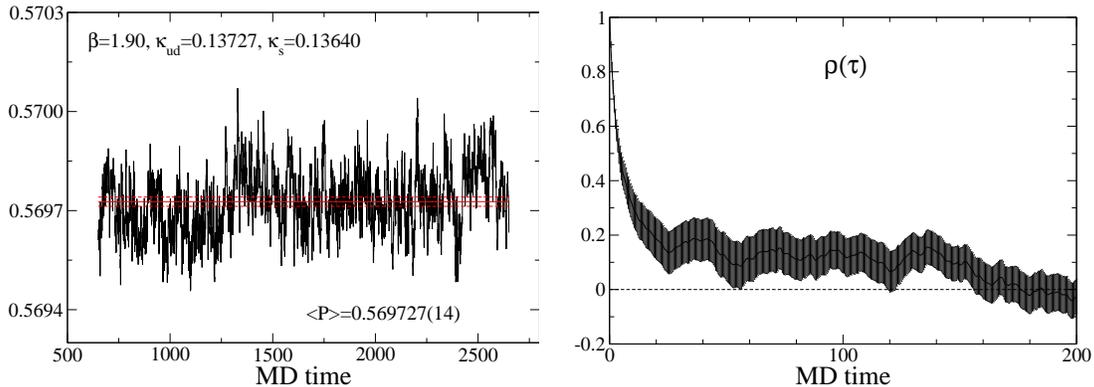

\vspace{3mm}
\begin{center}
\begin{tabular}{cc}
\includegraphics[width=71mm,angle=0]{figs/ukita/PLQ_his_b190kud013727ks013640b.eps}  &
\includegraphics[width=69mm,angle=0]{figs/ukita/PLQ_autocorr.eps} 
\end{tabular}
\end{center}
\vspace{-.5cm}
\caption{Plaquette history (left) and normalized autocorrelation 
function (right) for $(\kappa_{\rm ud}, \kappa_{\rm s})=(0.13727,
0.13640)$.  Horizontal lines in the left denote the average value of
the plaquette with an error band.}
\label{fig:PLQ}
\end{figure}

In Fig.~\ref{fig:PLQ} we show the plaquette history and the normalized 
autocorrelation function $\rho(\tau)$ at
$(\kappa_{\rm ud}, \kappa_{\rm s})=(0.13727, 0.13640)$ as a
representative case.
The integrated autocorrelation time 
is estimated as $\tau_{\rm int}[P]=20.9(10.2)$
following the definition in Ref.~\cite{luscher}  
\ben
\tau_{\rm int}(\tau)=\frac{1}{2}
+\sum_{0< \tau \le W} \rho(\tau),
\een
where the summation window $W$ is set to the first time lag $\tau$ that
$\rho(\tau)$ becomes consistent with zero within the error bar. 
In this case we find $W=119.5$. 
The choice of $W$ is not critical for estimate of 
$\tau_{\rm int}$ in spite of the long tail observed in Fig.~\ref{fig:PLQ}.
Extending the summation window, we find 
that $\tau_{\rm int}[P]$ saturates at $\tau_{\rm int}[P]\approx 25$ 
beyond $\tau=200$, which is
within the error bar of the original estimate.
For other hopping parameters 
we have found similar behaviors for the plaquette history and
the normalized autocorrelation function. 
We hardly observe the quark mass dependence for 
$\tau_{\rm int}[P]$ listed in Table~\ref{tab:param}. 
The statistics may not be
sufficiently large to derive a definite conclusion, however.

\section{Algorithmic issues}

\subsection{Efficiency of DDHMC algorithm}
\label{subsec:efficiency}

\begin{figure}[t!]
\vspace{3mm}
\begin{center}
\begin{tabular}{cc}
\includegraphics[width=75mm,angle=0]{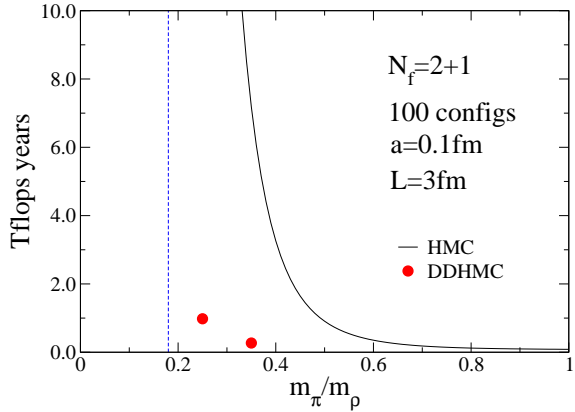}
\end{tabular}
\end{center}
\vspace{-.5cm}
\caption{Cost estimate of $N_f=2+1$ QCD simulations 
with the HMC (solid line) and the
DDHMC (red circle) algorithms at $a=0.1$~fm with $L=3$~fm 
for 100 independent configurations. Vertical dotted 
line denotes the physical point.}
\label{fig:berlinwall}
\end{figure}

In order to discuss the efficiency of the DDHMC algorithm,
it is instructive to compare with that of the HMC algorithm.
We first recall an empirical cost formula for
$N_f=2$ QCD simulations with the HMC algorithm 
based on the CP-PACS results\cite{berlinwall}:
\ben
{\rm cost[Tflops\cdot years]}&=&C\left[\frac{\#{\rm conf}}{1000}\right]\cdot
\left[\frac{0.6}{m_\pi/m_\rho}\right]^6\cdot
\left[\frac{L}{3{\rm ~fm}}\right]^5\cdot
\left[\frac{0.1{\rm ~fm}}{a}\right]^7\nn
\label{eq:cost_hmc}
\een
with $C\approx 2.8$.
A strong quark mass dependence is found in the above formula:
$1/(m_\pi/m_\rho)^6$ behaves as $1/m_{\rm ud}^3$ 
in the leading term for the small quark mass region. 
This quark mass dependence is owing to the following three factors:
The number of the quark matrix inversion is governed by the
condition number which should be proportional to $1/m_{\rm ud}$;
to keep the acceptance rate fixed we should take 
$\delta\tau\propto m_{\rm ud}$ for the step size in 
the molecular dynamics trajectories;
The autocorrelation time of the HMC evolution shows 
$1/m_{\rm ud}$ dependences in the CP-PACS run\cite{cppacs_nf2}.

To estimate the computational cost for
$N_f=2+1$ QCD simulations with the HMC algorithm,
we assume that the strange quark contribution is given by half of
eq.(\ref{eq:cost_hmc}) at $m_\pi/m_\rho=0.67$ which
is a phenomenologically estimated ratio of the
strange pseudoscalar meson ``$m_{\eta_{\rm s}}$'' and $m_\phi$:
\ben
\frac{m_{\eta_{\rm
s}}}{m_\phi}=\frac{\sqrt{2m_K^2-m_\pi^2}}{m_\phi}\approx 0.67.
\een
Since the strange quark is relatively heavy, its computational
cost occupies only a small fraction as the up-down quark masses decrease. 
In Fig.~\ref{fig:berlinwall} we draw the cost formula 
for the $N_f=2+1$ case as a function of $m_\pi/m_\rho$,
where we take \#conf=100, a=0.1~fm and $L=3$~fm in eq.(\ref{eq:cost_hmc})
as a representative case.
We observe a steep increase of the computational cost below 
$m_\pi/m_\rho\simeq 0.5$.
At the physical point the expected cost is $O(100)$ Tflops$\cdot$years.

Now let us turn to the case of the DDHMC algorithm.
The red symbol denotes the measured cost at 
$\kappa_{\rm ud}$=0.13781, 0.13770 with $\kappa_{\rm s}=0.13640$,
which are the lightest two points in our simulation.
The DDHMC algorithm show a remarkable improvement
reducing the cost by $30-50$ times in magnitude. 
The majority of this reduction arises from the multiple time scale
integration scheme and the GCR solver
accelerated by the SAP preconditioning
with the single-precision arithmetic.
Roughly speaking, the improvement factor is $O(10)$ for the former
and $3-4$ for the latter.
Note that the quark mass dependence is also tamed: 
Since we find that $\tau_{\rm int}[P]$ is independent 
of the quark masses, 
the cost is proportional to $1/m_{\rm ud}^2$.
Our results show a feasibility of simulations at the physical point
with the $O(10)$ Tflops computer which is available at present.

\subsection{$\tau$ dependence of DDHMC algorithm}

\begin{table}[t!]
\centering
\begin{tabular}{ccccccc}  \hline
 $\tau$  & $(N_0,N_1,N_2)$  &  $N_{\rm poly}$ & trajs. & MD time & 
$\tau_{\rm int}[P]$ & $\tau_{\rm int}[{\rm \#mult}]$\\ \hline
0.5   & (4,4,6) & 130 & 6000  & 3000 & 23.7(9.2) & 92(56) \\ 
0.5/3 & (4,4,2) & 130 & 18000 & 3000 & 18.6(5.9) & 42(21) \\ \hline
\end{tabular}
\caption{Parameters for $\tau$-dependence study. \#mult denotes
the number of multiplications
of the Wilson-Dirac quark matrix on the full lattice.}
\label{tab:tau_dep}
\end{table}

\begin{figure}[b!]
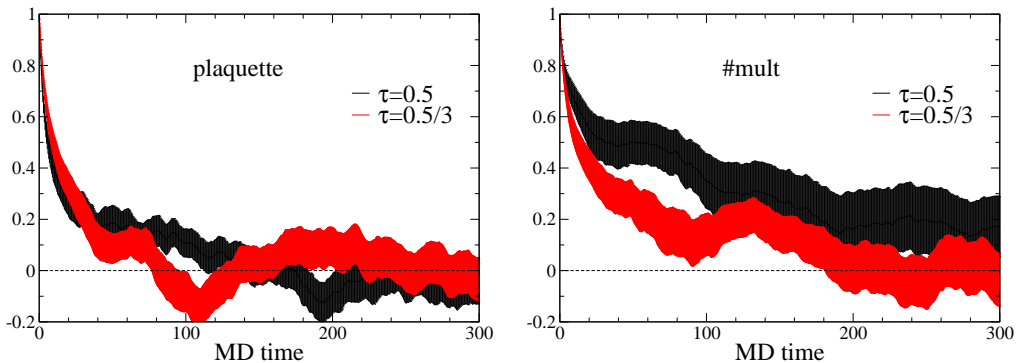

\vspace{3mm}
\begin{center}
\begin{tabular}{cc}
\includegraphics[width=65mm,angle=0]{figs/kura/plaq_ac_tau.eps} &
\includegraphics[width=65mm,angle=0]{figs/kura/mult_ac_tau.eps}
\end{tabular}
\end{center}
\vspace{-.5cm}
\caption{Normalized autocorrelation functions for the plaquette (left) 
and the \#mult (right) at $(\kappa_{\rm ud},\kappa_{\rm s})=(0.13700,0.13640)$.
Black (red) symbols denote the $\tau=0.5$ ($\tau=0.5/3$) case.}
\label{fig:tau_dep}
\end{figure}

In the DDHMC algorithm a subset of of all link variables,
which are referred to as the active link variables, are 
updated during the molecular dynamics evolution, while keeping
other field variables fixed\cite{luscher}. 
The fraction of the active link variables 
depends on the block size we choose. In our case of $8^4$ it is only
37\%. To ensure that all the link variables on the lattice should be 
updated equally on average we implement random gauge field 
translations at the end of every molecular dynamics trajectories
following Ref.~\cite{luscher}.

Our concern is that the DDHMC algorithm might have 
a long autocorrelation time due to the existence of fixed link variables 
during the molecular dynamics evolution.
A possible way to reduce the effects of the fixed link variables
is more frequent random gauge field translations.
This is easily realized by making $\tau$ shorter with
$\delta\tau$ fixed.
We investigate the $\tau$ dependence of the DDHMC algorithm
employing a smaller lattice size of $16^3\times 32$ at
$(\kappa_{\rm ud},\kappa_{\rm s})=(0.13700,0.13640)$.
Other parameters are summarized in Table~\ref{tab:tau_dep}.
 
In Fig.~\ref{fig:tau_dep} we show the normalized autocorrelation
functions for the plaquette and the number of multiplications
of the Wilson-Dirac quark matrix on the full lattice
during a molecular dynamics trajectory. 
Black symbols denote the $\tau=0.5$ case, while red ones are for
the $\tau=0.5/3$ case.
We observe that the normalized autocorrelation functions 
for $\tau=0.5$ have longer tails than $\tau=0.5/3$
before becoming consistent with zero:
This is quantitatively checked by the integrated autocorrelation times
$\tau_{\rm int}[P]$ and $\tau_{\rm int}[{\rm \#mult}]$
in Table~\ref{tab:tau_dep}: The $\tau=0.5/3$ case has
shorter autocorrelation times.
Since the computational cost is the same for
the $\tau=0.5$ and the $\tau=0.5/3$ cases in terms of MD time, 
we can conclude that 
the $\tau=0.5/3$ case shows a better efficiency than 
the $\tau=0.5$ case. 
Based on this study, albeit conducted at a relatively heavy quark mass
$(m_\pi/m_\rho\approx 0.6)$, we employ $\tau=0.25$
for the production run at $(\kappa_{\rm ud},\kappa_{\rm s})
=(0.13781,0.13640)$ and $(0.13770,0.13640)$, which is half
of the trajectory length at other hopping parameters.

\subsection{Simulation stability}

In Refs.~\cite{del06,del07} simulation stability was discussed
based on the spectral gap distribution of the Wilson-Dirac operator
for two-flavor lattice QCD simulations.
The spectral gap is defined as
\ben
\mu={\rm min}\{|\lambda|\; |\;
\lambda\;\mbox{is an eigenvalue of}\;Q\},
\een
where $Q$ is the hermitian Wilson-Dirac operator
$Q=\gamma_5 D_W$ with $D_W=(1/2)\{\gamma_\mu(\nabla_\mu^*+\nabla_\mu)
-a\nabla_\mu^*\nabla_\mu\}+m_0$.
Important indices to characterize the distribution are
its median ${\bar \mu}$ and width $\sigma$. 
The latter is defined as
$(v-u)/2$, where $[u,v]$ is the smallest range of $\mu$ which contains
more than 68.3\% of the data. This is to avoid potentially large
statistical uncertainties which might occur when data are not sufficiently
sampled.
Their chief findings are two points: The first one is that
the median ${\bar \mu}$ shows a good 
linear dependence on the current up-down quark mass $m_{\rm ud}^{\rm AWI}$
and the magnitude of the slope is well described by $Z_A$ empirically.
The second one is that the width $\sigma$ scales as
\ben
\sigma\frac{\sqrt{V}}{a}\simeq 1
\een
with $V$ the four-dimensional volume in physical units. 
They also observe that the width $\sigma$ is roughly independent of the
quark mass for the unimproved Wilson quark action, while it shows 
a trend to decrease with the mass for the improved one.

This study was also applied to the $N_f=2+1$ case 
in Ref.~\cite{kura_lat06}
where we reported on our preliminary run on a $16^3\times 32$ lattice 
preparing for the PACS-CS project.  
We observed ${\bar \mu}\propto m_{\rm ud}^{\rm AWI}$ and 
found $0.5\simlt \sigma({\sqrt{V}}/{a})\simlt 0.76$ for
$15{\rm ~MeV}<m_{\rm ud}^{\rm AWI}<64{\rm ~MeV}$, where
$\sigma$ diminishes as the up-down quark mass decreases.

The existence of a gap in the spectrum of the Wilson-Dirac operator
allows us to simulate the light quarks efficiently.
The authors in Ref.~\cite{del06} propose a stability condition requiring
${\bar \mu}\ge 3\sigma$ to assure the existence of the gap. 
Let us apply this condition to our case.
Assuming $\sigma({\sqrt{V}}/{a})= 1$ we estimate
$\sigma=2.26$~MeV using $a=0.09$~fm and $V=(2.8{\rm ~fm})^4$ which
will be obtained later.
By using the empirical relation 
${\bar \mu}\simeq Z_A m_{\rm ud}^{\rm AWI}$ 
we find $Z_A m_{\rm ud}^{\rm AWI}\simgt 6.8$~MeV for the
stability condition, which is heavier than the physical point.  
On the other hand, we found $\sigma({\sqrt{V}}/{a})< 1$ in 
Ref.~\cite{kura_lat06}, indicating that the actual bound will be lower.  
Our runs toward the physical point should shed light on the actual bound 
of stability for our lattice parameters.

\subsection{Comparison of DDHMC and mass-preconditioned HMC}

\begin{table}[t!]
\centering
\begin{tabular}{ccccccc}  \hline
prec. & block size  & $\rho$ & $(N_0,N_1,N_2)$  & $\tau$  & MD time & $P_{\rm acc}$ \\ \hline
DD   & $8^4$  & $-$  & (4,8,12) & 1 & 3000 & 0.857(8) \\ 
mass & $-$    & 0.09 & (4,8,12) & 1 & 3000 & 0.794(8) \\ \hline
\end{tabular}
\caption{Simulation parameters for the DDHMC 
and the mass-preconditioned HMC algorithms. $P_{\rm acc}$ denote the
acceptance rate.}
\label{tab:dd-mass}
\end{table}

\begin{figure}[b!]
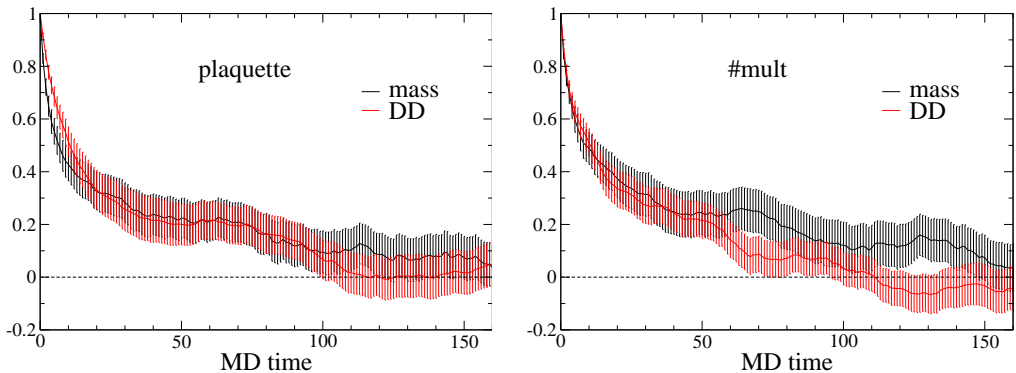

\vspace{3mm}
\begin{center}
\begin{tabular}{cc}
\includegraphics[width=65mm,angle=0]{figs/kura/plaq_ac_prec.eps} &
\includegraphics[width=65mm,angle=0]{figs/kura/mult_ac_prec.eps}
\end{tabular}
\end{center}
\vspace{-.5cm}
\caption{Normalized autocorrelation functions for the plaquette (left) 
and the number of multiplications of the Wilson-Dirac quark matrix on
the full lattice (right).
Black (red) symbols denote the DDHMC (mass-preconditioned HMC) case.}
\label{fig:dd-mass_autoc}
\end{figure}

As discussed in Sec.~\ref{subsec:efficiency}, 
it is essential for the efficiency of the DDHMC algorithm to incorporate
the multiple time scale integration scheme.  
It is well known that this scheme is also 
applicable to the mass-preconditioned HMC algorithm\cite{massprec1,massprec2}.
We have made a direct comparison of the two algorithms
in $N_f=2$ QCD on a $16^3\times 32$ lattice 
employing the $O(a)$-improved
Wilson quark action with the nonperturbative improvement coefficient
$c_{\rm SW}=2.0171$\cite{csw_nf2} 
and the plaquette gauge action at $\beta=5.2$.
The lattice spacing is 0.1~fm and the physical pseudoscalar meson mass
is about 600~MeV at $\kappa_{\rm ud}=0.1355$.
For the mass-preconditioned HMC algorithm we employ two set of 
the pseudofermion fields which decompose the fermion determinant as
\ben
\det Q^2=\det (W^\dagger W)\det\left(\frac{Q^2}{W^\dagger W}\right),
\een
where $Q$ is the hermitian Wilson-Dirac operator 
and the preconditioning operator is
given by $W=Q+\rho$.
For convenience we refer to $\det (W^\dagger W)$ as the UV part
and the $\det({Q^2}/(W^\dagger W))$ as the IR part in an analogy with
the DDHMC algorithm.
The step sizes are chosen with the
three integers $N_{0,1,2}$ in exactly the same way as the DDHMC algorithm.
Simulation parameters are summarized in Table~\ref{tab:dd-mass}.
The block size for the DDHMC algorithm and the $\rho$ parameter for the
mass-preconditioned HMC algorithm are chosen such that $||F_{0,1,2}||$
are roughly the same between these two algorithms.
This condition yields comparable acceptance ratios with $N_{0,1,2}$ in common. 
We employ the BiCGStab algorithm for the quark matrix inversion 
in both the UV and the IR parts.

\begin{table}[t!]
\centering
\begin{tabular}{cccccc}  \hline
prec. & $\tau_{\rm int}[P]$  & $\tau_{\rm int}[{\rm \#mult}]$  & 
\#mult  & cost[$P$]  & cost[\#mult]  \\ \hline
DD   & 27(10) & 22(7)  & 45530(280) & 1.2(5) & 1.0(3) \\ 
mass & 28(12) & 35(16) & 67160(380) & 1.9(8) & 2.4(1.0) \\ \hline
\end{tabular}
\caption{Integrated autocorrelation time and cost estimate 
for the DDHMC and the mass-preconditioned HMC algorithms.}
\label{tab:dd-mass_cost}
\end{table}

\begin{figure}[b!]
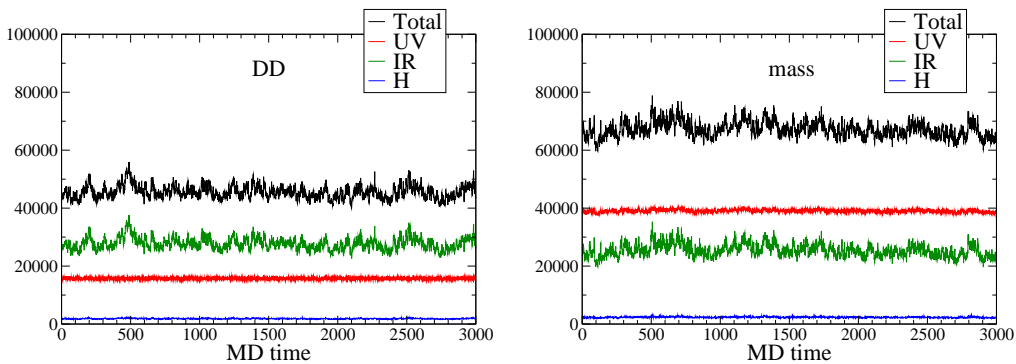

\vspace{3mm}
\begin{center}
\begin{tabular}{cc}
\includegraphics[width=65mm,angle=0]{figs/kura/mult_dd.eps} &
\includegraphics[width=65mm,angle=0]{figs/kura/mult_mass.eps}
\end{tabular}
\end{center}
\vspace{-.5cm}
\caption{History  
for number of multiplications of the Wilson-Dirac quark matrix 
on the full lattice for the DDHMC algorithm (left) 
and the mass-preconditioned HMC algorithm (right).}
\label{fig:dd-mass_hist}
\end{figure}

In Fig.~\ref{fig:dd-mass_autoc} we plot the normalized autocorrelation 
function for the plaquette as a function of MD time. 
The results of both algorithms show
quite similar behaviors and the integrated autocorrelation time  
$\tau_{\rm int}[P]$ in Table~\ref{tab:dd-mass_cost} 
are consistent within the errors.
Figure~\ref{fig:dd-mass_hist} shows the MD-time history of the 
number of multiplications of the Wilson-Dirac quark matrix 
on the full lattice.
The total number of multiplications is the sum of those required to
calculate the UV and the IR forces and the Hamiltonian.
Their contributions are denoted by black, red, green, blue lines in
order. 
Comparing the results of the DDHMC and the mass-preconditioned HMC,
we observe a clear difference in the UV part contribution:
the mass-preconditioned HMC needs more than twice of the
multiplication number for the DDHMC.  
This ends up in a 50\% difference in the total number of multiplications.   
In Fig.~\ref{fig:dd-mass_autoc} we also plot the normalized autocorrelation 
function for the total number of multiplications.
Although the DDHMC result seems to show a slightly steeper fall-off,
both results are consistent within the error bars.
This is confirmed by the integrated autocorrelation time 
$\tau_{\rm int}[{\rm \#mult}]$ in  Table~\ref{tab:dd-mass_cost}. 
  
Now let us compare the efficiencies of both algorithms.
We define the machine-independent cost formula by
\ben
{\rm cost}[{\cal O}]=\#{\rm mult(total)/MD\;\;time}\times 
\tau_{\rm int}[{\cal O}]/10^6, 
\een
where the observable $\cal O$ is the plaquette or the total number of
multiplications.
In Table~\ref{tab:dd-mass_cost} we summarize the results of
cost[${\cal O}$]. For both observables the DDHMC algorithm shows
better efficiency than the mass-preconditioned HMC algorithm
albeit the errors are rather large.

There remains a couple of concerns in this study.
The first one is the quark mass dependence,
because our results are obtained at only one hopping parameter.
The second one is the optimization.
While we choose $8^4$ block size for the DDHMC algorithm
and $\rho=0.09$ for the mass-preconditioned HMC algorithm since 
$||F_{0,1,2}||$ are roughly the same, these parameters may not be the optimal
values for each of the algorithms. We leave these issues to future studies.

\section{Physical results}

\subsection{Measurement of hadron masses, quark masses, decay constants}

\begin{figure}[t!]
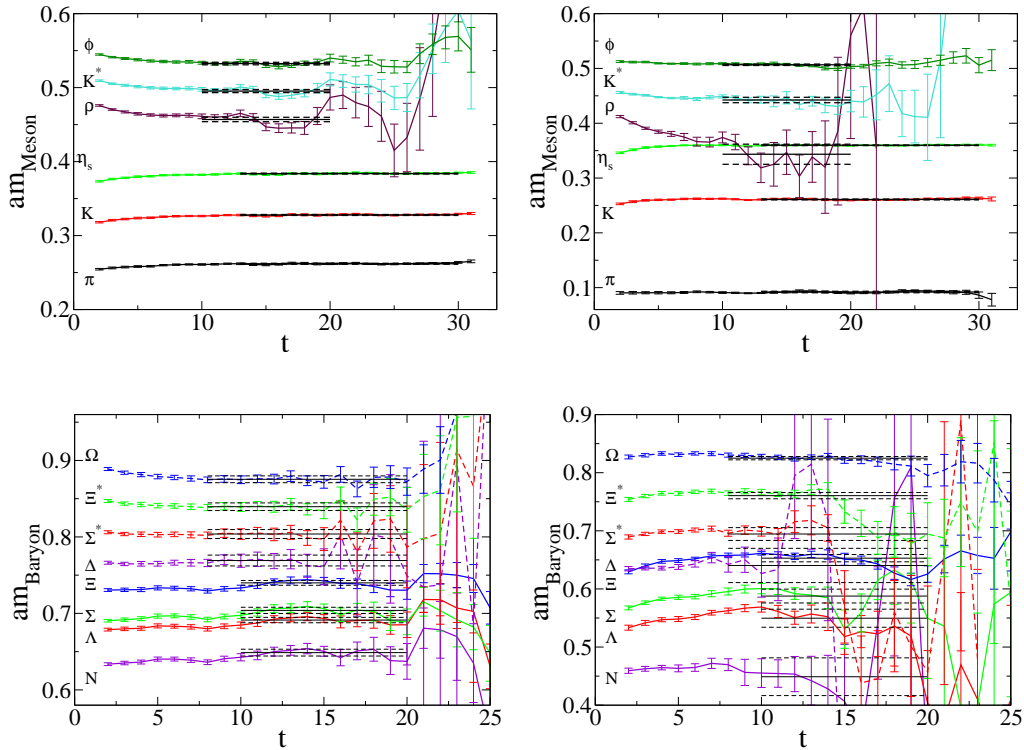

\vspace{3mm}
\begin{center}
\begin{tabular}{cc}
\includegraphics[width=65mm,angle=0]{figs/ukita/msn_13727.eps} &
\includegraphics[width=65mm,angle=0]{figs/ukita/msn_13781.eps} 
\vspace*{5mm}
\end{tabular}
\begin{tabular}{cc}
\includegraphics[width=65mm,angle=0]{figs/ukita/brn_13727.eps} &
\includegraphics[width=65mm,angle=0]{figs/ukita/brn_13781.eps}
\end{tabular}
\end{center}
\vspace{-.5cm}
\caption{Effective masses for the mesons (top) and the baryons
 (bottom) at $\kappa_{\rm ud}=0.13727$ (left) and 0.13781 (right).
Horizontal lines represent the fitting results with an error band.}
\label{fig:Meff}
\end{figure}

\begin{figure}[t!]
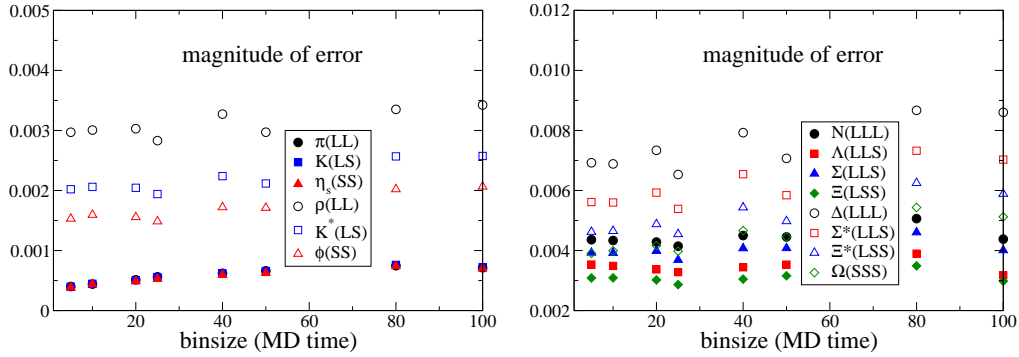

\vspace{3mm}
\begin{center}
\begin{tabular}{cc}
\includegraphics[width=65mm,angle=0]{figs/ukita/binsize_meson_13727.eps} &
\includegraphics[width=65mm,angle=0]{figs/ukita/binsize_baryon_13727.eps}
\end{tabular}
\end{center}
\vspace{-.5cm}
\caption{Binsize dependence of magnitude of error for mesons (left) and 
baryons (right) at $(\kappa_{\rm ud},\kappa_{\rm s})=(0.13727,0.13640)$.}
\label{fig:binsize}
\end{figure}

We measure both the meson and the baryon correlators at every 10 trajectories
at the unitary points where 
the valence quark masses are equal to the sea quark masses.
Light hadron masses are extracted from a single exponential 
$\chi^2$ fit to the correlators
with an exponentially smeared source and a local sink. 
Figure~\ref{fig:Meff} shows the hadron effective masses  
at $\kappa_{\rm ud}=0.13727$ and 
$0.13781$ as representative cases. 
We observe clear plateau for the mesons except for the $\rho$
meson and also good signal for the baryons thanks to a large volume.
Especially, the $\Omega$ baryon has a stable signal, which we use
as a physical input to determine the cutoff scale later.   
The horizontal lines denote the fitting results with an error band
of one standard deviation. Their widths represent the fitting ranges.
Statistical errors are estimated by the jackknife method.
In Fig.~\ref{fig:binsize} we plot binsize dependence of magnitude
of error for the mesons and the baryons at 
$(\kappa_{\rm ud},\kappa_{\rm s})=(0.13727,0.13640)$.
For the pseudoscalar mesons we observe that the
magnitude of error gradually increases as the bin size is enlarged up
to about 40 MD time, beyond which it stabilizes. 
For other hadrons we do not find any clear binsize dependence.
The data at other hopping parameters show similar behaviors.
Based on this observation
we choose 50 molecular dynamics time for the jackknife analysis
at all the hopping parameters.  

We extract the bare quark mass through the axial vector
Ward-Takahashi identity (AWI)  by 
\begin{eqnarray}
am^{\rm AWI}_q =\lim_{t\rightarrow \infty} \frac{\langle\nabla_4
  A_4^{\rm imp}(t) P(0)\rangle}{2\langle P(t)P(0)\rangle}
\end{eqnarray}
with $P$ the pseudoscalar density and $A_4^{\rm imp}$ 
the nonperturbatively $O(a)$-improved axial vector current\cite{ca}.
The renormalized quark mass and the
pseudoscalar meson decay constant in the continuum ${\overline{\rm MS}}$
scheme are defined as follows:
\begin{eqnarray}
m^{\overline{\rm MS}}_q&=&\frac{Z_A \left(1+b_A
  \frac{m^{\rm AWI}}{u_0}\right)}
{Z_P \left(1+b_P \frac{m^{\rm AWI}}{u_0} \right)}m^{\rm AWI}_q,\\
f_{\rm PS}&=&2\kappa u_0 Z_A \left(1+b_A\frac{m^{\rm AWI}_q}{u_0}\right) \frac{C_A^s}{C_P^s}\sqrt{\frac{2C_P^l}{m_{PS}}}.
\end{eqnarray}
Here $C_{A,P}^s$ are the amplitudes extracted from the correlators
$\langle A_4^{\rm imp}(t) P(0)\rangle$ and $\langle P(t)P(0)\rangle$
with an exponentially smeared source and a local sink, while
$C_P^l$ is from $\langle P(t)P(0)\rangle$ with a local source and
a local sink. The renormalization factors
$Z_{A,P}$ and the improvement coefficients $b_{A,P}$ are evaluated
perturbatively up to 
one-loop level\cite{z_pt,z_imp_pt}with the
tadpole improvement.

\subsection{Comparison with the previous CP-PACS/JLQCD results}

\begin{figure}[t!]
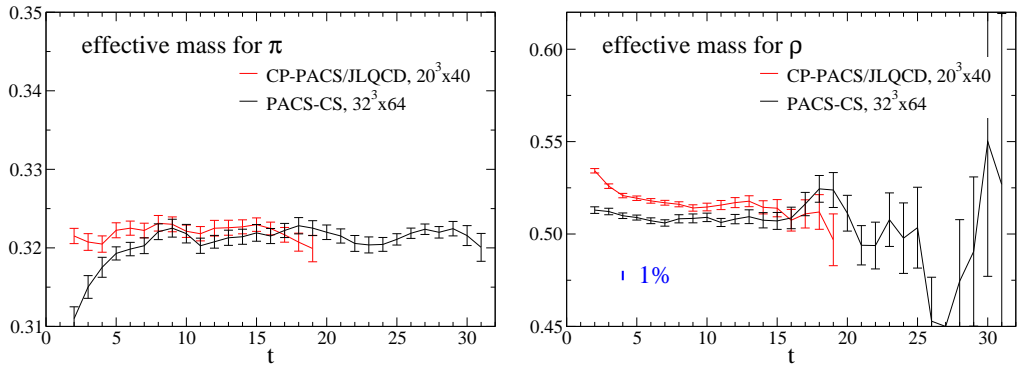

\vspace{3mm}
\begin{center}
\begin{tabular}{cc}
\includegraphics[width=65mm,angle=0]{figs/kura/e.msn_k22_s11_pi.eps} &
\includegraphics[width=65mm,angle=0]{figs/kura/e.msn_k22_s11_rho.eps}
\end{tabular}
\end{center}
\vspace{-.5cm}
\caption{Effective masses for the $\pi$ (left) and the $\rho$ (right) 
at $(\kappa_{\rm ud},\kappa_{\rm s})=(0.13700,0.13640)$.
Black and red symbols denote the PACS-CS and the CP-PACS/JLQCD results, 
respectively.}
\label{fig:comp_eff}
\end{figure}

\begin{table}[h!]
\centering
\begin{tabular}{ccccc}  \hline
                                 & lattice size           &  $am_{\pi}$ & $am_{\rho}$ & $am_{\rm N}$\\ \hline
PACS-CS     &$32^3\times 64$& 0.3220(6) & 0.506(2) & 0.726(3) \\
$[t_{\rm min},t_{\rm max}]$ & & [13,30] & [10,20] & [10,20] \\ 
CP-PACS/JLQCD&$20^3\times 40$ &0.3218(8)& 0.516(3) & 0.733(4)\\ 
$[t_{\rm min},t_{\rm max}]$ & & [8,20] & [9,15] & [11,17] \\ \hline
\end{tabular}
\caption{PACS-CS and CP-PACS/JLQCD results for 
$\pi$, $\rho$ and nucleon masses at 
$(\kappa_{\rm ud}, \kappa_{\rm s})=(0.13700, 0.13640)$.
[$t_{\rm min}$,$t_{\rm max}$] denotes the fitting range.}
\label{tab:comp}
\end{table}

\begin{figure}[b!]
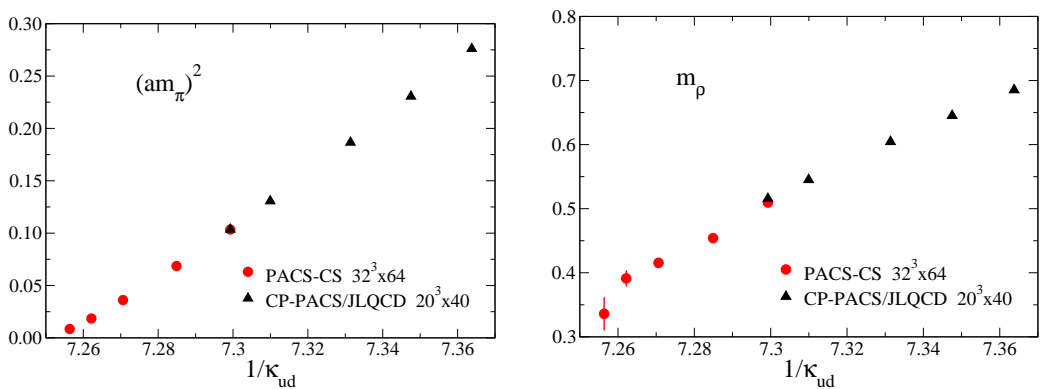

\vspace{3mm}
\begin{center}
\begin{tabular}{cc}
\includegraphics[width=65mm,angle=0]{figs/ukita/kud.PS_LL2.eps}
\hspace*{2mm} &
\includegraphics[width=65mm,angle=0]{figs/ukita/kud.V_LL.eps}
\end{tabular}
\end{center}
\vspace{-.5cm}
\caption{$(am_{\pi})^2$ (left) and $am_{\rho}$ (right) 
as a function of $1/\kappa_{\rm ud}$. 
Red and black symbols denote the PACS-CS and the CP-PACS/JLQCD results, 
respectively.}
\label{fig:comp}
\end{figure}

We first compare the PACS-CS results on $32^3\times 64$ 
with the previous CP-PACS/JLQCD results 
on $20^3\times 40$\cite{cppacs/jlqcd1, cppacs/jlqcd2} 
at $(\kappa_{\rm ud}, \kappa_{\rm s})=(0.13700, 0.13640)$.
In Fig.~\ref{fig:comp_eff} we plot the effective masses for the $\pi$
and the $\rho$ mesons. The PACS-CS and the CP-PACS/JLQCD
results are consistent for the $\pi$ meson, while a slight deviation
is observed for the $\rho$ meson. 
This is numerically confirmed by
the fitting results listed in Table~\ref{tab:comp}, where 
we employ a single exponential $\chi^2$ fit. 
The nucleon mass is also given in Table~\ref{tab:comp}.
We find 1$-$2\% deviation 
for the $\rho$ meson and nucleon masses, which
could be due to possible finite size effects. 

Figure~\ref{fig:comp} shows the up-down quark mass dependence of
$(am_{\pi})^2$ and $am_{\rho}$ with $\kappa_{\rm s}$ fixed at 0.13640.
For the pion mass we observe that the PACS-CS and 
the CP-PACS/JLQCD results are smoothly connected 
as a function of $1/\kappa_{\rm ud}$.
On the other hand, the quark mass dependence is not so smooth  
for the $\rho$ meson.
Although this may be attributed to finite size effects,
further studies are needed in the $\rho$ channel.

\subsection{Chiral analysis on pseudoscalar meson masses and decay constants}

We examine the chiral behaviors of the pseudoscalar meson masses
and decay constants in comparison with the prediction of chiral
perturbation theory (ChPT). Our interest exist in the following 
points: (i) signals for chiral logarithms, 
(ii) determination of low energy
constants in the chiral lagrangian, (iii) determination of the
physical point with the ChPT fit, (iv) estimate of the magnitude 
of finite size effects based on one-loop calculations of ChPT.

We first recall the one-loop expressions of ChPT 
for the pseudoscalar meson masses 
and the decay constants\cite{gasser}\footnote{
$f_\pi$ is normalized as 92.4~MeV in these expressions, while
our results are presented in the $f_\pi=130.7$~MeV normalization.}:
\begin{eqnarray}
\qquad 
m_{\pi}^2&=& 2 \hat m B_0 \left\{
1+\mu_\pi-\frac{1}{3}\mu_\eta 
+\frac{B_0}{f_0^2} \left(
16 \hat m  (2L_{8}-L_5) 
+16 (2 \hat m +m_{\rm s}) (2L_{6}-L_4)  
\right) \right\}, 
\label{eq:chpt_mpi}
\\
m_K^2&=&(\hat m +m_{\rm s})B_0 \left\{
1+\frac{2}{3}\mu_\eta
+\frac{B_0}{f_0^2}\left(
8(\hat m+m_{\rm s}) (2L_{8}-L_5)
+16(2 \hat m  +m_{\rm s}) (2L_{6}-L_4) 
 \right)\right\}, 
\label{eq:chpt_mk}
\\
f_\pi &=&f_0\left\{
1-2\mu_\pi-\mu_K
+ \frac{B_0}{f_0^2} \left (
8 \hat m L_5+8(2 \hat m +m_{\rm s})L_4
\right)\right\}, 
\label{eq:chpt_fpi}
\\
f_K &=&f_0\left\{
1-\frac{3}{4}\mu_\pi-\frac{3}{2}\mu_K-\frac{3}{4}\mu_\eta
+\frac{B_0}{f_0^2}\left(4(\hat m+m_{\rm s}) L_5+8(2 \hat m+ m_{\rm s})L_4 
\right)\right\}, 
\label{eq:chpt_fk}
\end{eqnarray}
where ${\hat m}=(m_{\rm u}+m_{\rm s})/2$ and 
$L_{4,5,6,8}$ are the low energy constants, and 
$\mu_{\rm PS}$ is the chiral logarithm defined by 
\begin{eqnarray}
\mu_{\rm PS}=\frac{1}{32\pi^2}\frac{m_{\rm PS}^2}{f_0^2}
\ln\left(\frac{m_{\rm PS}^2}{\mu^2}\right)
\label{eq:chlog}
\end{eqnarray}
with $\mu$ the renormalization scale.
There are six unknown low energy constants $B_0,f_0,L_{4,5,6,8}$ 
in the expressions above.
The low energy constants are scale-dependent so as to 
cancel that of the chiral logarithm (\ref{eq:chlog}). 
We determine these parameters                                  
by making a simultaneous fit for $m_\pi^2$, $m_K^2$, $f_\pi$ and $f_K$.

We also consider the contributions of the finite size effects based on ChPT.
At the one-loop level the finite size effects defined by 
$R_X=(X(L)-X(\infty))/X(\infty)$ for $X=m_\pi,m_K,f_\pi,f_K$ are given 
by~\cite{colangelo05}:
\begin{eqnarray}
&& \ \ \ R_{m_\pi}
=\frac{1}{4}\xi_{\pi}\tilde g_1(\lambda_\pi)-\frac{1}{12}\xi_{\eta}\tilde g_1(\lambda_\eta),\\
&& \ \ \ R_{m_K} 
=\frac{1}{6}\xi_{\eta}\tilde g_1(\lambda_\eta),\\
&& \ \ \ R_{f_\pi}
=-\xi_{\pi}\tilde g_1(\lambda_\pi)-\frac{1}{2}\xi_{K}\tilde g_1(\lambda_K),\\
&& \ \ \ R_{f_K} 
=-\frac{3}{8}\xi_{\pi}\tilde g_1(\lambda_\pi)-\frac{3}{4}\xi_{K}\tilde g_1(\lambda_K)
-\frac{3}{8}\xi_{\eta}\tilde g_1(\lambda_\eta)
\end{eqnarray}
with
\begin{eqnarray}
\hspace{2.5cm} && \xi_{\rm PS} \equiv \frac{m_{\rm PS}^2}{(4\pi f_\pi)^2},
\quad \ \lambda_{\rm PS} \equiv m_{\rm PS} L, \quad
\tilde g_1(x)=\sum_{n=1}^{\infty}\frac{4m(n)}{{\sqrt n }x} K_1({\sqrt n} x), 
\end{eqnarray}
where $K_1$ is the Bessel function of the second kind and $m(n)$
denotes the multiplicities in the expression of $n=n_x^2+n_y^2+n_z^2$. 
With the use of these formulae
we estimate the possible finite size effects in our results.

\begin{figure}[t!]
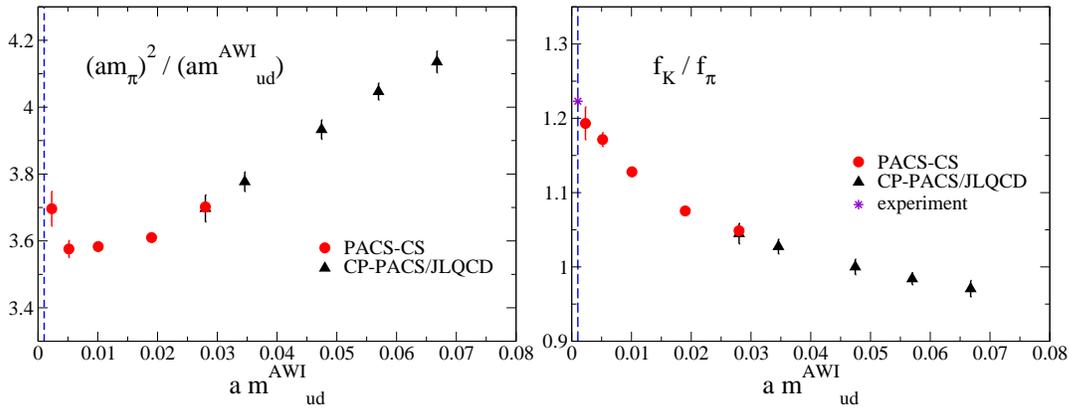

\begin{center}
\hspace{-0.6cm}
\includegraphics[width=7cm,keepaspectratio,clip]{figs/kadoh/f.mpi2bmud.cp.eps}
\includegraphics[width=7cm,keepaspectratio,clip]{figs/kadoh/f.fKfpi.cp.eps}
\caption{Comparison of the PACS-CS (red) and the CP-PACS/JLQCD (black)
  results for $(am_\pi)^2/(am_{\rm ud}^{\rm AWI})$ (left) and
  $f_K/f_\pi$ (right) as a function of $am_{\rm ud}^{\rm AWI}$. 
$\kappa_{\rm s}$ is fixed at 0.13640.
Vertical lines denote the physical point and star symbol represents
the experimental value.}
\label{fig:comparison}
\end{center}
\end{figure}

Before presenting our fitting results, it is instructive to
compare the PACS-CS and the CP-PACS/JLQCD results for 
$(a m_\pi)^2/(am_{\rm ud}^{\rm AWI})$ and $f_K/f_\pi$.
In Fig.~\ref{fig:comparison} we plot them 
as a function of $am_{\rm ud}^{\rm AWI}$ with
$\kappa_{\rm s}$ fixed at 0.13640. The PACS-CS and the CP-PACS/JLQCD
results are denoted by the red and the black symbols, respectively.
The two sets of data together show a smooth behavior as 
a function of $am_{\rm ud}^{\rm AWI}$, and at $\kappa_{\rm ud}=0.13700$  
($am_{\rm ud}^{\rm AWI}=0.028$) they show good consistency.
It is important to observe that an almost linear quark mass dependence of the 
CP-PACS/JLQCD results for heavier up-down quark masses changes into a convex behavior,  
both for $(a m_\pi)^2/(am_{\rm ud}^{\rm AWI})$ and $f_K/f_\pi$, 
as the quark mass is lowered in the PACS-CS runs. 
This is a characteristic feature expected from 
the ChPT prediction in the small quark mass region
due to the chiral logarithm.
This curvature drives up the ratio $f_K/f_\pi$ toward the 
experimental value as the physical point is approached.

\begin{table}[b!]
\begin{center}
\hspace{1cm}
\begin{tabular}{ccccc}\hline
$\kappa_{\rm ud}$ & $\kappa_{\rm s}$  & $am_{\pi}$  & $am_{\rm ud}^{\rm AWI}$ & 
$am_{\rm s}^{\rm AWI}$ \\
\hline
\hline
0.13700  &  0.13640  &  0.32196(62)  &  0.02800(20) & 0.04295(30)  \\
0.13727  &  0.13640  &  0.26190(66)  &  0.01895(13) & 0.04061(18)  \\
0.13754  &  0.13640  &  0.18998(56)  &  0.01020(11) & 0.03876(18)  \\ 
         &  0.13660  &  0.17934(78)  &  0.00908(7)  & 0.03257(17)  \\
0.13770  &  0.13640  &  0.13591(88)  &  0.00521(9)  & 0.03767(10)  \\
0.13781  &  0.13640  &  0.08989(291) &  0.00227(16) & 0.03716(20)  \\
\hline
\end{tabular}
\caption{Pion masses and unrenormalized AWI quark masses.}
\label{tab:quarkmass}
\end{center}
\end{table}

\begin{figure}[t!]
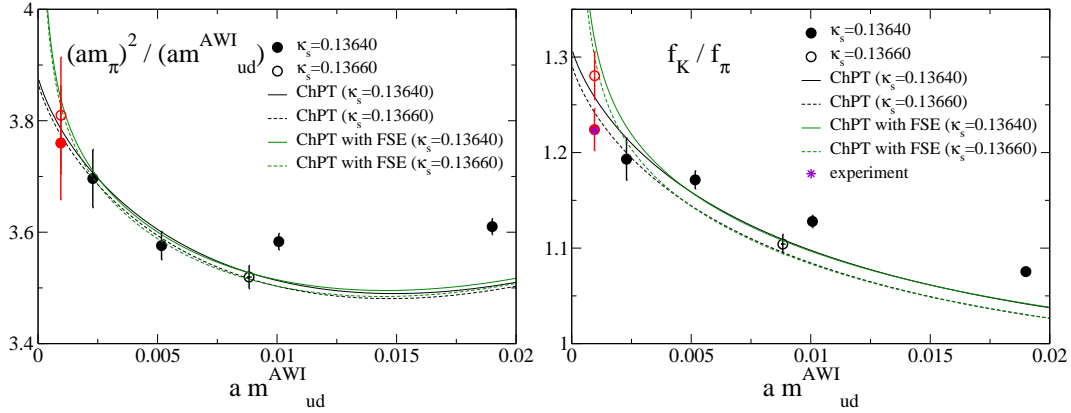

\begin{center}
\hspace{-0.6cm}
\includegraphics[width=7cm,keepaspectratio,clip]{figs/kadoh/f.mpi2bmud.eps}
\includegraphics[width=7cm,keepaspectratio,clip]{figs/kadoh/f.fKfpi.eps}
\caption{Fitting results for $(am_\pi)^2/(am_{\rm ud}^{\rm AWI})$
  (left) and $f_K/f_\pi$ (right). Red solid (open) symbols denote the
 extrapolated values at the physical point by the ChPT formulae
without (with) the finite size effects. }
\label{fig:fit}
\end{center}
\end{figure}

Let us apply the ChPT formulae (\ref{eq:chpt_mpi})$-$(\ref{eq:chpt_fk})
to our results at four points 
$(\kappa_{\rm ud},\kappa_{\rm s})=(0.13781,0.13640)$,
(0.13770,0.13640), (0.13754,0.13640), (0.13754,0.13660).  For these points, 
the $\rho$ meson mass satisfies the condition that $m_\rho > 2m_\pi$.
The measured bare AWI quark masses 
are used for ${\hat m}$ and $m_{\rm s}$ 
in eqs.(\ref{eq:chpt_mpi})$-$(\ref{eq:chpt_fk}).
The heaviest pion mass at $(\kappa_{\rm ud},\kappa_{\rm s})=(0.13754,0.13640)$
is about 430~MeV with the use of the cutoff determined below.
We summarize the pion masses and the unrenormalized AWI quark masses
in Table~\ref{tab:quarkmass}.
The fit results are shown in Fig.~\ref{fig:fit}, 
where the black solid lines
are drawn with $\kappa_{\rm s}$ fixed at 0.13640 and the black dotted 
lines are for $\kappa_{\rm s}=0.13660$. 
The red solid symbols represent the extrapolated values at the
physical point whose determination is explained 
in Sec.~\ref{sec:physicalpt} below. 
The heaviest point at 
$(\kappa_{\rm ud},\kappa_{\rm s})=(0.13754,0.13640)$
is not well described by ChPT both for 
$(a m_\pi)^2/(a m_{\rm ud}^{\rm  AWI})$ and $f_K/f_\pi$, and 
$\chi^2$/d.o.f. is rather large (see  Table~\ref{tab:fit}).

\begin{table}[h!]
\centering
\begin{tabular}{ccccc}\hline
$L_i(\mu=m_\eta)$ &PACS-CS  & PACS-CS with FSE      & exp. value\cite{amoros01}  
& MILC\cite{bernard06}
    \\
\hline
\hline
$L_4$        & 0.25(11)  &  0.23(12)      & 0.27 $\pm$ 0.8  & 0.1(2)(2)           \\
$L_5$        &  2.28(13) &  2.29(14)      & 2.28 $\pm$ 0.1  & 2.0(3)(2)           \\
$2L_6-L_4$ & 0.16(4)   &  0.16(4)          & 0 $\pm$ 1.0     & 0.5(1)(2)    \\ 
$2L_8-L5$  & $-$0.59(5) & $-$0.60(5)         & 0.18 $\pm$ 0.5  & $-$0.1(1)(1)    \\
\hline
$\chi^2$/d.o.f. & 2.1(1.4) &  2.1(1.4) & &  \\
\hline
\end{tabular}
\caption{Results for the low energy constants together with the
phenomenological estimates\cite{amoros01} and the MILC results\cite{bernard06}.}
\label{tab:fit}
\end{table}

The results for the low energy constants are presented
in Table~\ref{tab:fit} where the
phenomenological values with the experimental
inputs\cite{amoros01} and the MILC results\cite{bernard06}
are also given for comparison.
The renormalization scale is chosen to be $m_\eta=0.547$~GeV.   
For $L_4$ and $L_5$ governing the behavior of $f_\pi, f_K$, 
our results show good agreement with both the
phenomenological estimates and the MILC results. 
On the other hand, some discrepancies are observed between three results 
for $2L_6-L_4$ and $2L_8-L_5$ which enter into the ChPT formulae 
for $m_\pi^2$ and $m_K^2$.

\begin{figure}[t!]
\vspace{3mm}
\begin{center}
\begin{tabular}{cc}
  \includegraphics[width=65mm]{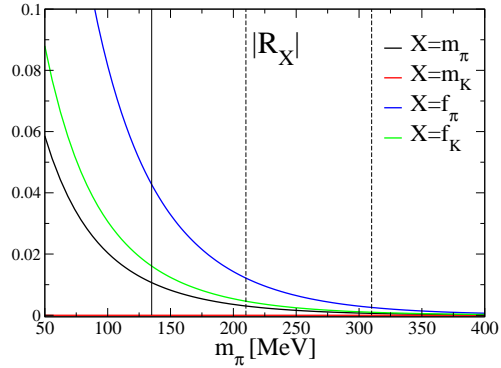}
\end{tabular}
\end{center}
\vspace{-.5cm}
\caption{$|R_X|$ ($R_{m_{\rm PS}}>0$ and $R_{f_{\rm
      PS}}<0$) for $X=m_\pi,m_K,f_\pi,f_K$ with $L=2.8$~fm 
as a function of $m_\pi$.
Solid vertical line denotes the physical point and the dotted ones
are for our simulation points.}
\label{fig:ratio}
\end{figure}

In Fig.~\ref{fig:fit} we also draw the ChPT fit results 
including the finite size effects. The green solid
lines are drawn for $\kappa_{\rm s}=0.13640$ and the green dotted ones
for $\kappa_{\rm s}=0.13660$. The fit curves with and without 
the finite size effects are almost degenerate for 
$a m_{\rm ud}^{\rm  AWI}>0.003$, but deviations appear 
closer to the physical point,  for which the extrapolated values are 
plotted by the open and solid red symbols.
This feature is understood by Fig.~\ref{fig:ratio} where
we plot the magnitude of $R_X$ for $X=m_\pi,m_K,f_\pi,f_K$ 
with $L=2.8$~fm as a function of $m_\pi$
( we note that $R_{m_{\rm PS}}>0$ and $R_{f_{\rm PS}}<0$).  
The finite size effects are less than 2\% for 
$m_{\rm PS}$ and $f_{\rm PS}$ at our simulation points.    
For $m_{\rm PS}$ this is true even at the physical point, while 
for $f_\pi$ the finite size effects cause the value to decrease by 4\%.

%
%


\subsection{Physical point and light hadron spectrum}
\label{sec:physicalpt}

\begin{figure}[b!]
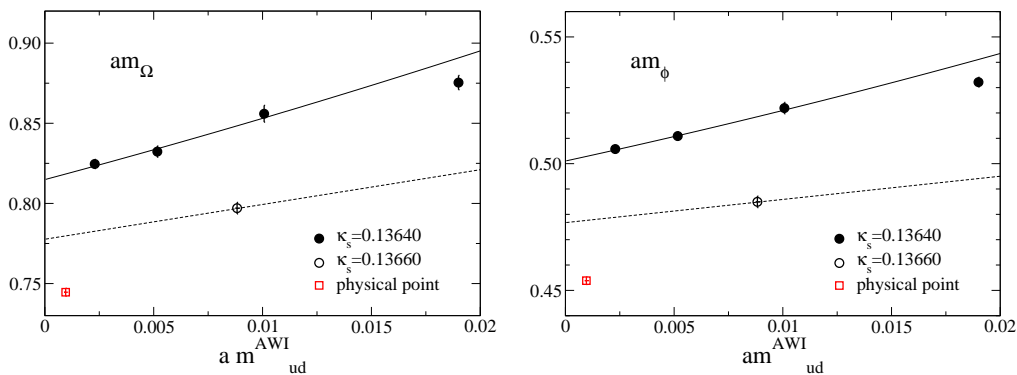

\vspace{3mm}
\begin{center}
\begin{tabular}{cc}
\includegraphics[width=65mm,angle=0]{figs/kadoh/f.Del_SSS.eps} &
\includegraphics[width=65mm,angle=0]{figs/kadoh/f.V_SS.eps}
\end{tabular}
\end{center}
\vspace{-.5cm}
\caption{Linear chiral extrapolation for $am_\Omega$ (left) 
and $am_{\phi}$ (right). 
Solid (dotted) lines are drawn with $\kappa_{\rm s}=0.13640$
(0.13660).
Red open symbols denote the extrapolated values at the physical point
with a linear form.}
\label{fig:phiomega}
\end{figure}

In order to determine the up-down and the strange quark masses and the
lattice cutoff we need three physical inputs.
We try the following two cases: 
$m_\pi, m_K, m_\Omega$ and $m_\pi, m_K, m_\phi$. 
The choice of $m_\Omega$ has theoretical and practical advantages:
the $\Omega$ baryon is stable in the strong interactions
and its mass, being composed of three strange quarks, 
is determined with good precision with small finite 
size effects. 
We also choose $m_\phi$ for comparison.
We employ the NLO ChPT formulae for the chiral extrapolations
of $m_\pi$, $m_K$, $f_\pi$ and $f_K$.  A simple linear formula
$m_{\rm had}=c_0+c_1\cdot m_{\rm ud}^{\rm AWI}+c_2\cdot m_{\rm s}^{\rm AWI}$ is
used for the other hadron masses, employing data in the same range   
$\kappa_{\rm ud}\ge 0.13754$ as for the pseudoscalar mesons.
In Fig.~\ref{fig:phiomega} we show the linear chiral extrapolations
for $m_\phi$ and $m_\Omega$. The solid lines are drawn with 
$\kappa_{\rm s}$ fixed at 0.13640 and the dotted ones are for 
$\kappa_{\rm s}=0.13660$. 
We observe that the quark mass dependences for $m_\phi$ and $m_\Omega$ 
at $\kappa_{\rm ud}\ge 0.13754$ are well described by the linear function.

\begin{table}[t!]
\centering
\begin{tabular}{ccccccc}  \hline
  input  & $a^{-1}$~[GeV]  &  $m^{\overline{\rm MS}}_{ud}$~[MeV] & 
$m^{\overline{\rm MS}}_{s}$~[MeV] & $f_\pi$ & $f_K$ & $f_K/f_\pi$\\ \hline
$m_\Omega$ & 2.256(81)  & 2.37(11) & 69.1(25) & 144(6) & 175(6) & 1.219(22) \\
$m_\phi$   & 2.248(76)  & 2.38(11) & 69.4(25) & 143(6) & 175(5) & 1.219(21) \\
\hline
\end{tabular}
\caption{Cutoff, renormalized 
quark masses, pseudoscalar meson decay constants determined with 
$m_\Omega$ and $m_\phi$ inputs.}
\label{tab:phypoint}
\end{table}

The results for the quark masses and the lattice cutoff are listed in
Table~\ref{tab:phypoint},
where the errors are statistical. The two sets of results are
consistent within the error.  The quark masses
are smaller than the recent estimates in the literature.  We note, however, 
that we employed the perturbative renormalization factors to one-loop level 
which may contain  a sizable uncertainty.  A nonperutrbative calculation 
of the renormalization factor is in progress using the Schr\"odinger 
functional scheme. 

In Table~\ref{tab:phypoint} we also present
predictions for the pseudoscalar meson decay constants
at the physical point
using the physical quark masses and the cutoff determined above, 
which should be compared with the experimental values 
$f_\pi=130.7$~MeV, $f_K=159.8$~MeV, $f_K/f_\pi = 1.223$.
A 10\% discrepancy in the magnitude of $f_\pi$ and $f_K$ 
might be due to use of one-loop perturbative $Z_A$ 
since the ratio shows a good agreement.  
A nonperturbative calculation of $Z_A$ 
is also in progress.

\begin{figure}[t!]
\vspace{3mm}
\begin{center}
\begin{tabular}{cc}
\includegraphics[width=70mm,angle=0]{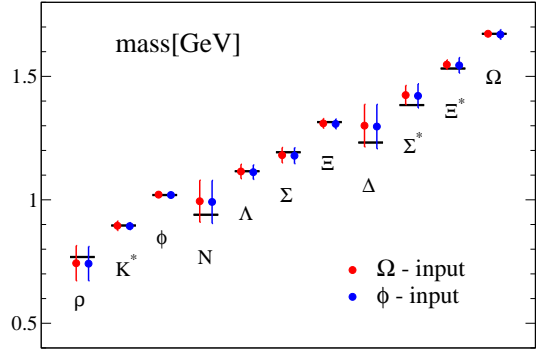}
\end{tabular}
\end{center}
\vspace{-.5cm}
\caption{Light hadron spectrum extrapolated at the physical point 
with  $\Omega$-input (red) and $\phi$-input (blue).
Horizontal bars denote the experimental values.}
\label{fig:spectrum}
\end{figure}


In Fig.~\ref{fig:spectrum} we compare the light hadron spectrum 
extrapolated to the physical point with the experiment.
The results for the $\Omega$-input and the $\phi$-input  
are consistent with each other, and both are in agreement 
with the experiment albeit errors are still not small 
for some of the hadrons. This is an encouraging result.  
However, further work is needed since 
cutoff errors of $O((a\Lambda_{\rm QCD})^2)$ are present 
in our results.

\section{$\rho$-$\pi\pi$ mixing}
\begin{figure}[t!]
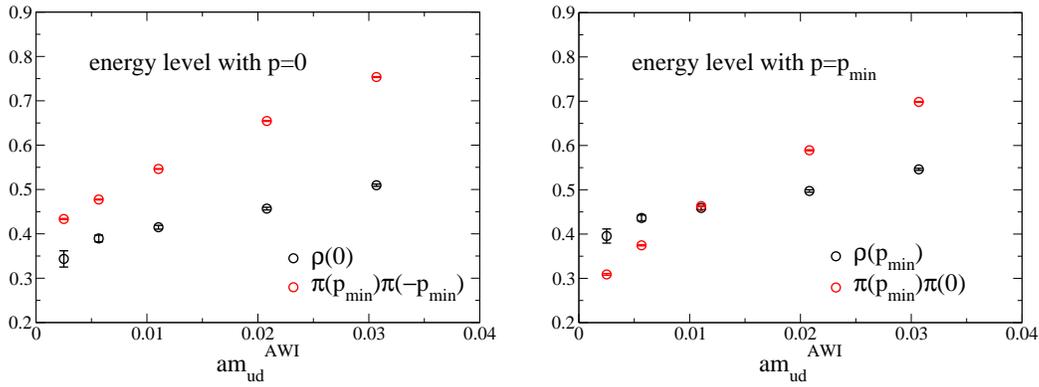

\vspace{3mm}
\begin{center}
\begin{tabular}{cc}
\includegraphics[width=65mm,angle=0]{figs/kura/E-level_p0.eps}  
\hspace*{2mm}&
\includegraphics[width=65mm,angle=0]{figs/kura/E-level_p1.eps}
\end{tabular}
\end{center}
\vspace{-.5cm}
\caption{Energy levels of the $\rho$ meson and the two pion states
without the total momentum $p=0$ (left) and 
with $p=p_{\rm min}\equiv 2\pi/L$ (right) as a function of the
up-down quark mass. $\kappa_{\rm s}$ is fixed at 0.13640.}
\label{fig:rhopipi_energy}
\end{figure}

\begin{figure}[t!]
\vspace{3mm}
\begin{center}
\begin{tabular}{cc}
\includegraphics[width=55mm,angle=0]{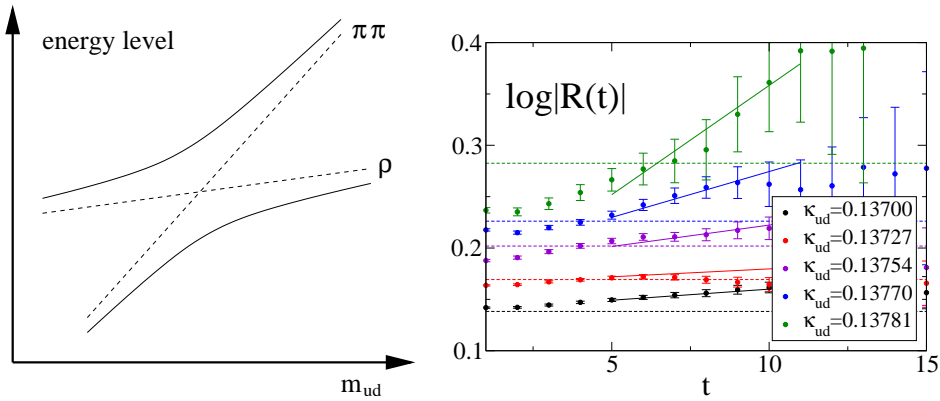}  &
\includegraphics[width=65mm,angle=0]{figs/ukita/log.msn_p1_k1_s11_rho_cpca_sy.eps}
\end{tabular}
\end{center}
\vspace{-.5cm}
\caption{Schematic view of $\rho$ and $\pi\pi$ energy levels 
due to mixing effects (left)
and time dependence of the $R$ function (right).}
\label{fig:rhopipi_r}
\end{figure}

Since our simulations are carried out at sufficiently 
small up-down quark masses, 
it would be interesting to investigate the $\rho$-$\pi\pi$ mixing effects.
We find that the rest mass $m_\rho$ is always smaller than 
the two-pion energy $2\sqrt{m_\pi^2+(2\pi/L)^2}$ for all the hopping parameters,
and hence the $\rho$ meson at rest cannot decay into two pions. 
However, as illustrated in Fig.~\ref{fig:rhopipi_energy}, for 
a moving $\rho$ with a unit of momentum, {\it i.e.,}, its energy  
$\sqrt{m_\rho^2+(2\pi/L)^2}$ becomes larger than the energy of a moving 
pion and a pion at rest given by 
$\sqrt{m_\pi^2+(2\pi/L)^2}+m_\pi$ when the up-down quark mass is sufficiently reduced. 

Let us consider two types of the $\rho$ meson propagator with the
momentum $2\pi/L$: $\rho_\|(2\pi/L)$ with polarization parallel
to the spatial momentum and $\rho_\bot(2\pi/L)$ with polarization 
perpendicular to the spatial momentum.
Phenomenologically the $\rho$-$\pi\pi$ coupling is described by
$g_{\rho\pi\pi}\epsilon_{abc}\rho_\mu^a\pi^b\partial_\mu\pi^c$,
which favors $\rho_\|(2\pi/L)\rightarrow \pi(2\pi/L)\pi(0)$
to $\rho_\bot(2\pi/L)\rightarrow \pi(2\pi/L)\pi(0)$.
We expect that the $\rho_\|(2\pi/L)$ propagator is more
strongly affected by the mixing effects than
the $\rho_\bot(2\pi/L)$ correlator.
Since the mixing effects push up the upper energy level further and
push down the lower energy level as shown in Fig.~\ref{fig:rhopipi_r},
they could be detected by measuring the $R$ function defined by
\begin{eqnarray}
R(t)=\frac{\langle\rho_\|({\vec p},t)\rho_\|^\dagger({\vec p},0)\rangle}
{\langle\rho_\bot({\vec p},t)\rho_\bot^\dagger({\vec p},0)\rangle}
\stackrel{{\rm large\;\;}t}{\longrightarrow}
Z{\rm e}^{-(E_{\rho_\|}-E_{\rho_\bot})t}.
\label{eq:ratio}
\end{eqnarray} 
In Fig.~\ref{fig:rhopipi_r} we plot $\log|R(t)|$ as a function of $t$.
The dotted horizontal lines denote $R(t)=(E/m_\rho)^2$ , which is determined
kinematically in the mixing-free case.
The solid lines represent the fitting
results with a single exponential form over $5\le t\le 11$.
The data show clear positive slopes which indicate 
$E_{\rho_\|}<E_{\rho_\bot}$. 
We also observe that the magnitude of the
energy difference is rather small for $\kappa_{\rm ud}\le 0.13754$, while
it grows rapidly as the up-down quark mass is reduced for  
$\kappa_{\rm ud} > 0.13754$.
This feature may suggest that the 
$\langle\rho_\|({\vec p},t)\rho_\|^\dagger({\vec p},0)\rangle$ 
correlator is getting dominated by the $\pi\pi$ state toward
the smaller up-down quark masses.
In order to obtain a definite conclusion, we 
need more detailed investigations with increased statistics.

\section{Summary}
We have presented a status report of the PACS-CS project which aims at 
a 2+1 flavor lattice QCD simulation toward the physical point. 
With the aid of the DDHMC algorithm for the up-down quarks we have 
reached $m_{\pi}=210$~MeV,  which roughly corresponds to 
$m_{\rm ud}^{\overline{\rm MS}}(\mu=2{\rm ~GeV})=5.6$~MeV, 
on a $32^3\times 64$ lattice using the $O(a)$-improved Wilson quarks. 
Thanks to the enlarged volume 
compared to the previous CP-PACS/JLQCD work,
we obtain good signals not only for the meson masses 
but also for the baryon masses. 
Our results for the hadron spectrum at the physical point
show a good agreement with the experimental values.
  
At present we have just started the simulation at the physical point.
We are also calculating  the nonperturbative renormalization factors
for the quark masses and the pseudoscalar meson decay constants in 
order to remove perturbative uncertainties in these important quantities.
Once these calculations are accomplished,
the next step is to investigate the finite size effects 
at the physical point,  and then to reduce the discretization errors by 
carrying out calculations at finer lattice spacings. 

\vspace{5mm}
\noindent
{\bf \large Acknowledgment}

We would like to thank all the collaboration members for discussions
and, in particular, A.~Ukawa for a careful reading of this report.
Numerical calculations for the present work have been carried out
under the ``Interdisciplinary Computational Science Program'' in
Center for Computational Sciences, University of Tsukuba.
This work is supported in part by Grants-in-Aid for Scientific Research
from the Ministry of Education, Culture, Sports, Science and Technology
(Nos.~13135204, 15540251, 17340066, 17540259, 18104005, 18540250, 18740139).

\end{document}